\documentclass[12pt,preprint]{aastex}

\slugcomment{To appear in AJ}

\shorttitle{A search for binary black holes in AGN}
\shortauthors{Tingay \& Wayth}

\begin{document}


\title{A VLBA search for binary black holes in active galactic nuclei with double-peaked optical emission line spectra}


\author{S.J. Tingay}
\affil{International Centre for Radio Astronomy Research, Curtin University, Bentley WA, Australia}
\email{s.tingay@curtin.edu.au}

\author{R. Wayth}
\affil{International Centre for Radio Astronomy Research, Curtin University, Bentley WA, Australia}



\begin{abstract}
We have examined a subset of 11 active galactic nuclei (AGN) drawn from a sample of 87 objects that possess double-peaked optical emission line spectra, as put forward by \citet{wan09a} and are detectable in the FIRST survey at radio wavelengths.  The double-peaked nature of the optical emission line spectra has been suggested as evidence for the existence of binary black holes in these AGN, although this interpretation is controversial.  We make a simple suggestion, that direct evidence of binary black holes in these objects could be searched for in the form of dual sources of compact radio emission associated with the AGN.  To explore this idea, we have used the Very Long Baseline Array to observe these 11 objects from the \citet{wan09a} sample.  Of the 11 objects, we detect compact radio emission from two, SDSS J151709$+$335324 and SDSS J160024$+$264035.  Both objects show single components of compact radio emission.  The morphology of SDSS J151709$+$335324 is consistent with a recent comprehensive multi-wavelength study of this object by \citet{ros10}.  
Assuming that the entire sample consists of binary black holes, we would expect of order one double radio core to be detected, based on radio wavelength detection rates from FIRST and VLBI surveys. We have not detected any double cores, thus this work does not substantially support the idea that AGN with double-peaked optical emission lines contain binary black holes.  However, the study of larger samples should be undertaken to provide a more secure statistical result, given the estimated detection rates.
\end{abstract}


\section{Introduction}
A study of a sample of 87 active galactic nuclei (AGN) that display double-peaked optical emission line
spectra, by \citet{wan09a,wan09b}, shows evidence for correlations between the ratios of
the shifts of the red shifted and blue shifted emission lines and the respective line strengths. \citet{wan09a}
interpret these data as due to the existence of binary supermassive black holes in these
AGN, the result of hierarchical minor and/or major merging, as predicted by $\Lambda$CDM cosmology.
\citet{wan09a} suggest that the typical separation of the dual black holes is approximately 1
kpc, representing an intermediate stage of merging systems.

The optical emission line data presented are striking in many cases and the suggestion that the AGN may harbour binary black holes
warrants further investigation using independent approaches. \citet{wan09a} cite some independent evidence that the
sources in their sample contain binary black holes, from the occurrence of one X-shaped radio
source in their sample of 87 AGN (plus another X-shaped radio source with a similar optical host not in their sample). X-shaped radio sources have been suggested to be the result of
binary black hole and accretion disk systems launching two sets relativistic jets from the nuclear
region of AGN \citep{mer02}, along different trajectories.

A far better direct tracer of binary black holes separated by ~1 kpc in these AGN could be dual compact
radio cores, as revealed by high-resolution radio imaging. Not only would the compact cores
directly indicate the existence of black holes, they would allow the measurement of the separation
of the black holes easily, further testing the suggestions of \citet{wan09a}.
Recently, other searches for AGN with double-peaked emission lines have been made, based on
methods very similar to those used by \citet{wan09a}, by \citet{smi09} and by \citet{liu10}, all based on the Sloan Digital Sky Survey (SDSS) Data Release 7 (DR7).  However, very few studies to date have undertaken spatially resolved radio observations of these samples.  The notable exception is the comprehensive study of SDSS~J151709$+$335324 (an object that features in this paper) by \citet{ros10}.  \citet{liu10} note the need for spatially resolved imaging at radio, X-ray and optical wavelengths, for samples of AGN with double-peaked optical emission line spectra.

Other previous work by \citet{bor09} claims SDSS 153636.22+044127.0 is a binary black hole system with separation of 0.1 pc.  \citet{gas10} advances an alternative model of double-peaked optical line emission from a disk.  Also, \citet{liu10} and \citet{ros10} discuss an explanation in terms of AGN outflows or jets.  Follow-up imaging and spectroscopy of the \citet{liu10} sample by \citet{she11} favours narrow-line region kinematics associated with single AGN systems as the explanation for the emission line profiles.  Thus, multiple mechanisms for double-peaked optical emission lines appear to be plausible.  A VLBA investigation of this class of AGN using a direct search for binary black holes traced by dual compact radio cores is therefore a valuable test that can be undertaken in a straightforward manner.  Determining the incidence of supermassive binary black hole systems is of interest due to their role in producing gravitational radiation that can potentially be detected by existing pulsar timing arrays and future gravity wave detectors.

In section 2 we describe the selection criteria and the resulting sample of radio sources, the VLBA observations that were undertaken, and the results of the observations.  In section 3 we discuss these results within the context of a binary black hole interpretation for these objects and discuss the individual objects.

\section{Sample Selection, Observations and results}

We defined a set of radio sources selected from the \citet{wan09a} sample of 87 double-peaked optical emission
line AGN from the SDSS DR7 database. When we cross-correlated the positions of the \citet{wan09a} sample
sources with the objects in the FIRST database (flux density threshold for catalog of 1 mJy) \citep{bec95}, we found that 11 of the sample objects were detected
in FIRST, selecting those objects listed with radio emission within 5\arcsec~ of the SDSS
position. The 5\arcsec~ positional matching radius is generous, as it is of order 10 times the FIRST positional accuracy.  We found that the 11 sources with positional matches at 5\arcsec~ tolerance would have been matched with a much tigher tolerance of 1\arcsec, as all matches were 0.8\arcsec~ or better.  The 11 sources are listed in Table 1.  It is worth noting that the X-shaped radio source in the \citet{wan09a} sample does not appear in Table 1 because of a 26\arcsec~ offset between the listed FIRST position and the SDSS position.  Thus, this source does not meet our positional matching criterion.  This individual object likely warrants VLBI follow-up in its own right.

Observations of these 11 sources were undertaken with the VLBA over a 24 hour period on 2010 June 6 at a frequency of 1.4 GHz (observation code BT108).  Nine of the ten VLBA antennas were available for this experiment (the Pie Town antenna was not available).  The observations consisted of eight minute scans of the target sources, interspersed with two minute scans of nearby phase reference sources.  Switching angles between target and reference pairs ranged between 0.6$^{\circ}$ and 2.4$^{\circ}$.  Each target and reference were observed while above the horizon at all VLBA stations.  Therefore, over the 24 hour period, between one and two hours of integration time were obtained for each of the 11 targets.  A typical $(u,v)$ coverage is shown in Figure 1.

The observations were made with dual circularly polarised bands consisting of 4$\times$16 MHz IFs.  Correlation took place using the DiFX software correlator in Socorro \citep{del07}, producing visibilities in 32$\times$0.5 MHz channels per IF at a correlator integration period of 2 seconds.  At this time and frequency resolution, the accessible field of view for imaging extends far beyond the extent of the host galaxies.

The correlated data were analysed initially in AIPS, using the standard calibration routines described in the AIPS Cookbook.  The visibility amplitudes were calibrated using the $T_{sys}$ and gain values recorded at the VLBA stations.  The phase reference sources were fringe-fitted and the solutions applied to the target source data.  The resulting calibrated visibilities were written to disk as FITS files, for further imaging in DIFMAP \citep{she95}.

The visibility data were inspected, to note any strong detections.  Only one detection was noted, for SDSS~J151709$+$335324.  From this strong detection, we could assess the impact of the time and angle switching between targets and phase reference calibrators, via consideration of self-calibration corrections for this object.  For SDSS~J151709$+$335324, with a switching angle of 1.2$^{\circ}$ to its reference source, we found that the self-calibrated phases produced a dirty image with peak brightness 10\% higher than produced before self-calibration.  Thus, while some loss of coherence was inflicted because of time and angle switching, it is minimal in terms of the overall sensitivity of the observations.

In DIFMAP, images were produced over a 2\arcsec~ field, centred on the FIRST position, using 4096 pixel images with pixel size of approximately 1 mas.  This field is approximately 8 times larger than the accuracy of the FIRST positions, easily adequate to detect compact emission if centred on the brightest components of the FIRST sources.  The RMS image noises were typically 0.5 mJy/beam, giving a 3$\sigma$ detection threshold of 1.5 mJy/beam (enough sensitivity to detect the weakest source in the sample if the full flux density was from compact components).  It should be noted that in a relatively shallow VLBA survey such as reported here, the possibility exists to mis-identify core-jet, GHz-Peaked Spectrum or Compact Steep Spectrum radio sources as double core AGN.  Any double core AGN detected here would need to be verified using further deep observations that determine the detailed structure of the source and the spectral indices of the double compact components.

Two of the sources were detected in imaging, SDSS~J151709$+$335324 and SDSS~J160024$+$264035.  SDSS~J160024$+$264035 consists of a single weak point source approimately 0.\arcsec 1 away from the listed FIRST position, with a flux density of 1.8 mJy.  An image of SDSS~J160024$+$264035 appears in Figure 2a.  SDSS~J151709$+$335324 also has a single component morphology, of 17 mJy, shown in Figure 2b.  Errors on the measured flux densities are approximately 10\%, dominated by the uncertainty in the flux density scale.

Larger images were made for all objects, over a field of 4\arcsec, to search for any strong extended emission.  None was detected.

\section{Discussion}

\subsection{Double radio cores and binary black holes in AGN with double-peaked optical emission line spectra}

These observations have been made on the suggestion that binary black holes in AGN with double-peaked optical emission line spectra may be revealed as double radio cores in high resolution VLBI images.

From 87 double-peaked AGN, 11 (13\%) were found to have radio emission at 1.4 GHz from the FIRST survey, above the detection threshold of 1 mJy, with optical/radio position matches better than 5\arcsec.  Previously, \citet{bes05} found that the existence of optical emission lines is not a predictor of radio emission, rather the radio detected fraction of galaxies is a strong function of stellar mass, with fractions ranging from almost zero to 30\%, with higher fractions corresponding to higher stellar masses, considering objects with radio luminosities at 1.4 GHz of $>10^{23}$ W/Hz.  The corresponding radio luminosities for our sample typically lie between $10^{22}$ and $10^{24}$ W/Hz, not identically matched to the \citet{bes05} criterion, meaning that the prediction of detection fractions for our work based on the \citet{bes05} work is somewhat uncertain.  In other previous work, \citet{lee10} show that among Seyfert and LINER galaxies in SDSS, the radio detected fraction ranges between approximately 2\% and 10\%, depending on the particular optical sub-class and whether the FIRST or NVSS radio survey was used.  Thus, the 13\% radio detected fraction of the 87 double-peaked AGN does not appear unusual in terms of the radio detected fraction of the broader classes of emission line AGN, with radio detected fraction likely more dependent on stellar mass than emission line properties.

Of the 11 sources detected in FIRST, compact radio emission was detected in two, approximately 18\%.  Previous work by \citet{gar05} on a deep VLBI survey of the NOAO Bootes field shows that the detection rate of sub-mJy radio sources at 1.4 GHz is 8$^{+4}_{-5}$\% and for mJy radio sources is 29$^{+11}_{-12}$\%.  In other VLBI surveys of FIRST sources, \citet{por04} found VLBI detection rates of 33\% and \citet{wro04} finds 60\%.  The detection rate from our VLBI observations appears broadly consistent with this ensemble of previous work.

Thus, the sample of 87 AGN does not appear to be special in terms of its radio properties.  The incidence of radio emission and the incidence of compact radio emission appear similar to other samples previously studied.

Neither of the two objects detected have morphologies that unambiguously allow identification of binary black holes.  Both are simple weak point sources, easily explained as emission from a single radio core, presumably relatively weak radio emission coincident with a black hole.  Any companion black holes would have to have radio emission below the detection limit in our images, or be substantially further away than 4\arcsec.

Thus, we conclude that none of the 87 double-peaked AGN from \citet{wan09a} contain an incidence of double radio cores at our detection sensitivity that would support the existence of binary black holes in these AGN.  

To estimate the expected detection rate of binary black holes in our sample, we proceed as follows.  First, we assume that all 87 objects in the emission line sample host binary black holes.  This is the most aggresive assumption possible and allows no other physical mechanism that produces similar optical emission line profiles.  Given the radio detection rates from \citet{bes05} and \citet{lee10} listed above, a detection probability for radio emission of between 0\% and 30\% can be adopted.  Further, the detection rates for compact radio structure from VLBI observations, given the existence of radio emission, is variously estimated to be between 10\% and 60\%, as described above.  Given these probabilities and assuming that the probability of detectable compact radio emission is independent and identical for each of the two black holes in a binary system, the probability that both binary black holes will be detected as compact radio emission, for any given object from the parent sample is:

\[P(BBH)=P(RE).P(CPE|RE)^{2}\],

where $P(BBH)$ is the probability of detecting both binary black holes as compact radio sources, $P(RE)$ is the probability that an object has detectable radio emission in FIRST, and $P(CRE|RE)$ is the probability that an object has compact radio emission, given that it is detectable in FIRST.  The second probability is squared due to being independent for each black hole. Thus, taking all the conservative assumptions and maximal estimates of these two probabilities ($P(RE)=0.3$ and $P(CRE|RE)=0.6$) gives $P(BBH)\sim0.1$.  Taking mid-range estimates ($P(RE)=0.1$ and $P(CRE|RE)=0.3$) gives $P(BBH)\sim0.01$.  Taking low estimates for $P(RE)$ and $P(CRE|RE)$ gives $P(BBH)<<0.01$.  It is worth noting that even taking the maximal estimates of the probabilities will give a lower detection rate if the aggresive assumption that all double-peaked optical emission line galaxies is relaxed.

Adopting the mid-range estimates for the detection probabilities appears a reasonable approach and gives $P(BBH)\sim0.01$.  Thus, under this assumption, at most one binary black hole could be expected from our current observations, using the calculations above and the physical interpretation of \citet{wan09a}.

Clearly the study of larger samples is required and future work will include similar observations of the samples of \citet{liu10} and \citet{smi09} 

\subsection{The nature of the radio emission in SDSS~J151709$+$335324}

While no double radio cores or evidence for binary black holes were found in this sample, SDSS~J151709$+$335324 has been the subject of a recent multi-wavelength study by \citet{ros10}.  From FIRST, SDSS~J151709$+$335324 has a flux density of 120 mJy at 1.4 GHz \citep{bec95} with the FIRST image of the object showing that it consists of a single unresolved component at the angular resolution of the survey ($\sim$5\arcsec).  The VLBA observations reported here reveal a single compact component, totalling approximately 17 mJy, $\sim$13\% of that seen in FIRST.  SDSS~J151709$+$335324 has a reported redshift of 0.135, obtained from the SDSS data.  This redshift corresponds to a luminosity distance of 615 Mpc (based on H$_{o}=73$ km/s/Mpc, $\Omega_{matter}=0.27$ and $\Omega_{vacuum}=0.73$).  

Recently, \citet{ros10} presented a comprehensive multi-wavelength study of SDSS~J151709$+$335324, including multi-frequency VLA images.  \citet{ros10} put forward two plausible explanations for their muti-wavelength data: a) a multiple-AGN produced due to a massive dry merger or b) a very powerful radio jet driven outflow that produces the different velocity components seen in optical spectra.  Rosario et al. favour the second explanation, on the basis that their multi-frequency VLA images show prominent jet-like features and a compact core that remains bright at 22 GHz. 

\citet{ros10} calculates an astrometric position for the radio source of RA=15:17:09.24228; DEC=$+$33:53:24.5663 (J2000), with an error of 25 mas.  The single compact component detected with the VLBA is located at RA=15:17:09; DEC=$+$33:53:24.5 (J2000) in our image, with an estimated 3-$\sigma$ error of approximately 120 mas.  This error is entirely dominated by the uncertainty in the position of the phase reference calibrator used for this observation, which is approximately 40 mas according to the VLBA calibrator list\footnote{http://www.vlba.nrao.edu/astro/calib/}, drawing from the best known position \citep{bro98}.  Thus, the astrometric position of \citet{ros10} coincides with the VLBA position of the brightest compact component to within the positional errors.  

Under the favoured interpretation of \citet{ros10}, the compact radio component detected with the VLBA would represent the core and the origin of the jet that provides the energy to power the optical emission lines and the kinematic variations observed in the optical spectra.  

We note alternative explanations for SDSS~J151709$+$335324.  An examination of other FIRST sources within 1.5 arcmin of SDSS~J151709$+$335324 reveal nearby radio sources which, when taken together with SDSS~J151709$+$335324, could be described as a classic FR-II radio galaxy morphology (Figure 3).  If this is the case, SDSS~J151709$+$335324 would be associated with one of the two hot spots of this radio galaxy (Figure 3) and the VLA images of Rosario et al. (2010) could be interpreted as describing a double hot spot structure.  Such an interpretation admits a variation on the conclusions of \citet{ros10}.  Instead of the emission lines regions driven by a jet originating in SDSS~J151709$+$335324, the emission line region could be powered by a jet that impinges on SDSS~J151709$+$335324 from a distant origin.  Examples of this are known, for example in PKS 0344$-$345 \citep{len09}.

However, we note that under this interpretation, no host galaxy for the FR-II object can be identified in SDSS.  We also note that while the compact VLBA component bears a resemblance to compact components seen in other FR-II hot spots, such as in Pictor A \citep{tin08}, the luminosity of the compact components in SDSS~J151709$+$335324 would have to be 10,000 times greater than those seen in Pictor A.  For these two reasons, we find that this interpretation would appear to be unlikely.  A further alternative scenario is that the VLA and VLBA emission arises from a dual AGN system, with one AGN revealing compact radio emission above our detection threshold (VLBA detection), but the second AGN having no such bright compact emission (VLA centroid of emission).  This alternative could be tested with deeper VLBA observations that seek to detect compact emission coincident with the VLA emission centroid, in additon to the compact emission detected in the observations presented here.

Further dedicated follow-up observations of SDSS~J151709$+$335324 with the VLBA will be useful to provide better sensitivity and image quality for this object, and are being planned, to possibly detect the jet emission near the core.  Based on our data, interpreted in conjunction with the comprehensive data of Rosario et al. (2010), we conclude that the simplest explanation for this object is that the compact radio emission is related to the core of the galaxy.

\section{Conclusion}

We have imaged, using the VLBA, 11 of 87 AGN with double-peaked optical emission line spectra and radio emission, suggested by \citet{wan09a} to harbour binary black holes.  We used the images to search for double radio cores in these objects, as a direct tracer of binary black holes.  The incidence of radio emission and compact radio emission in this sample appears typical and in agreement with previous studies.  No evidence of double radio cores was found from our observations.  Thus, no evidence for binary black holes was found.  However, it is likely that, given the small probability of detection (under various assumptions), larger samples of double-peaked emission line galaxies will need to be studied with VLBI in order to make a useful statistical test.

{\it Facilities:} \facility{VLBA}.

\clearpage

\begin{figure}
\plotone{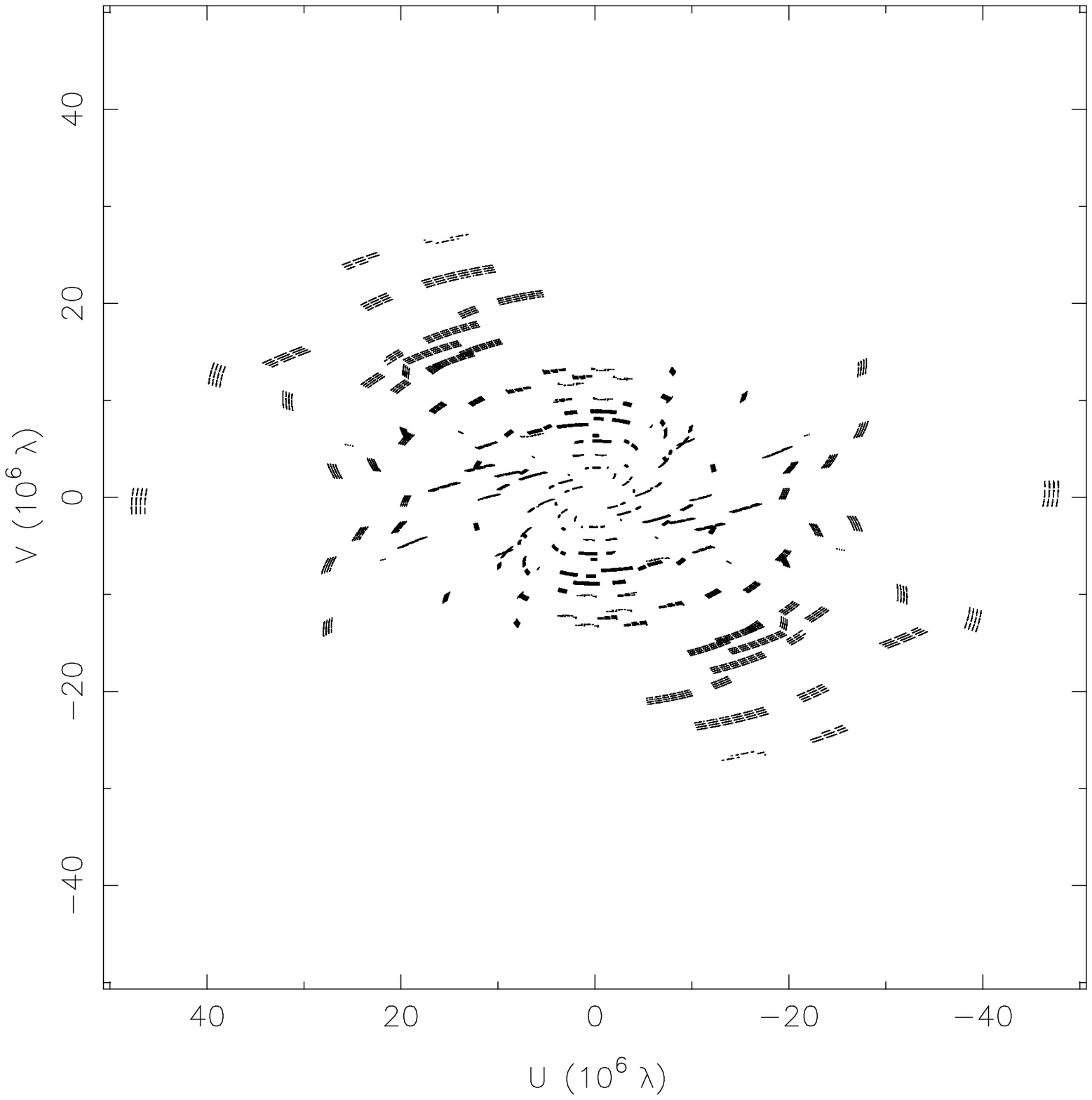}
\caption{Typical $(u,v)$ coverage for the VLBA observations.}
\end{figure}

\clearpage

\begin{figure}
\plottwo{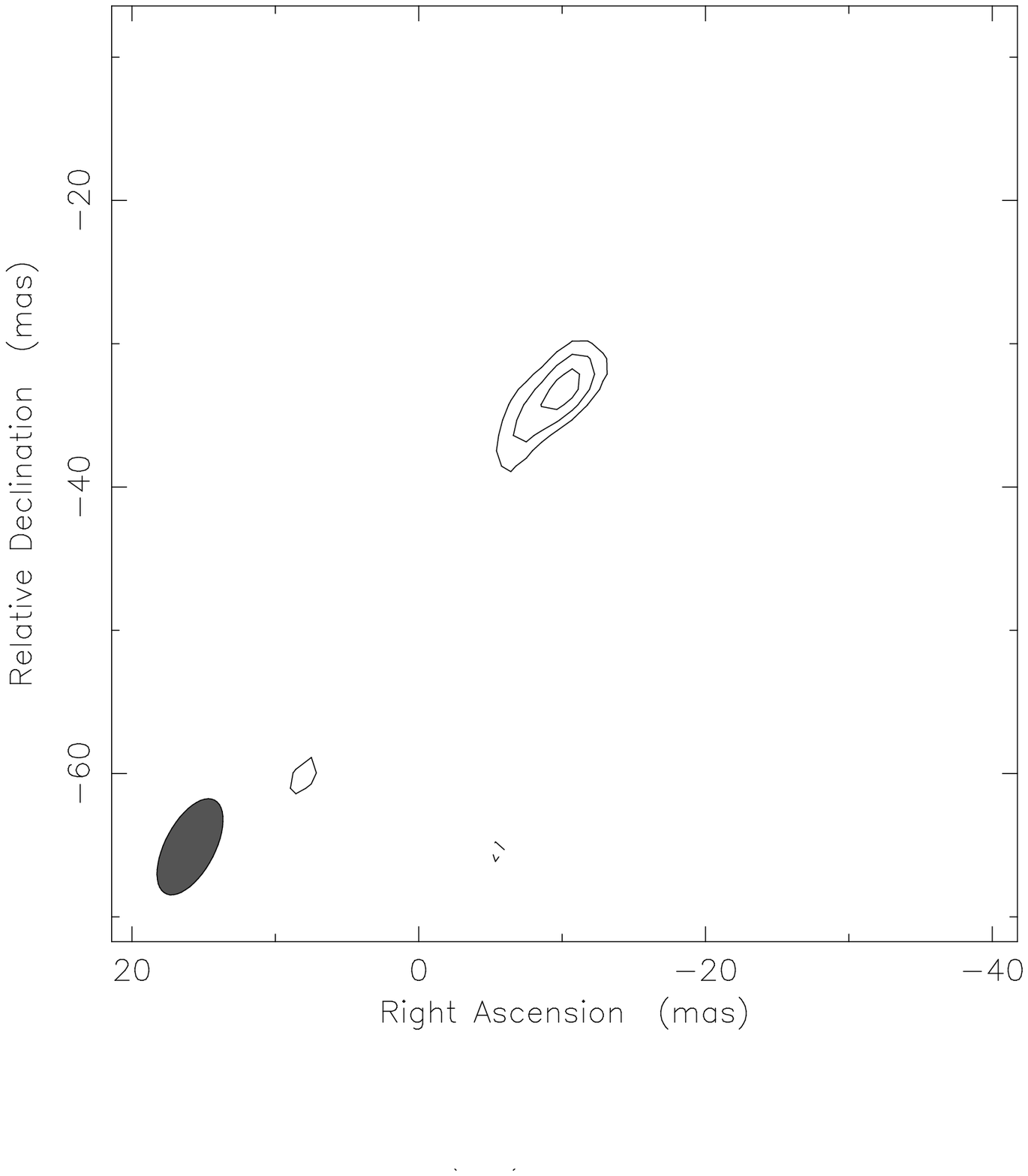}{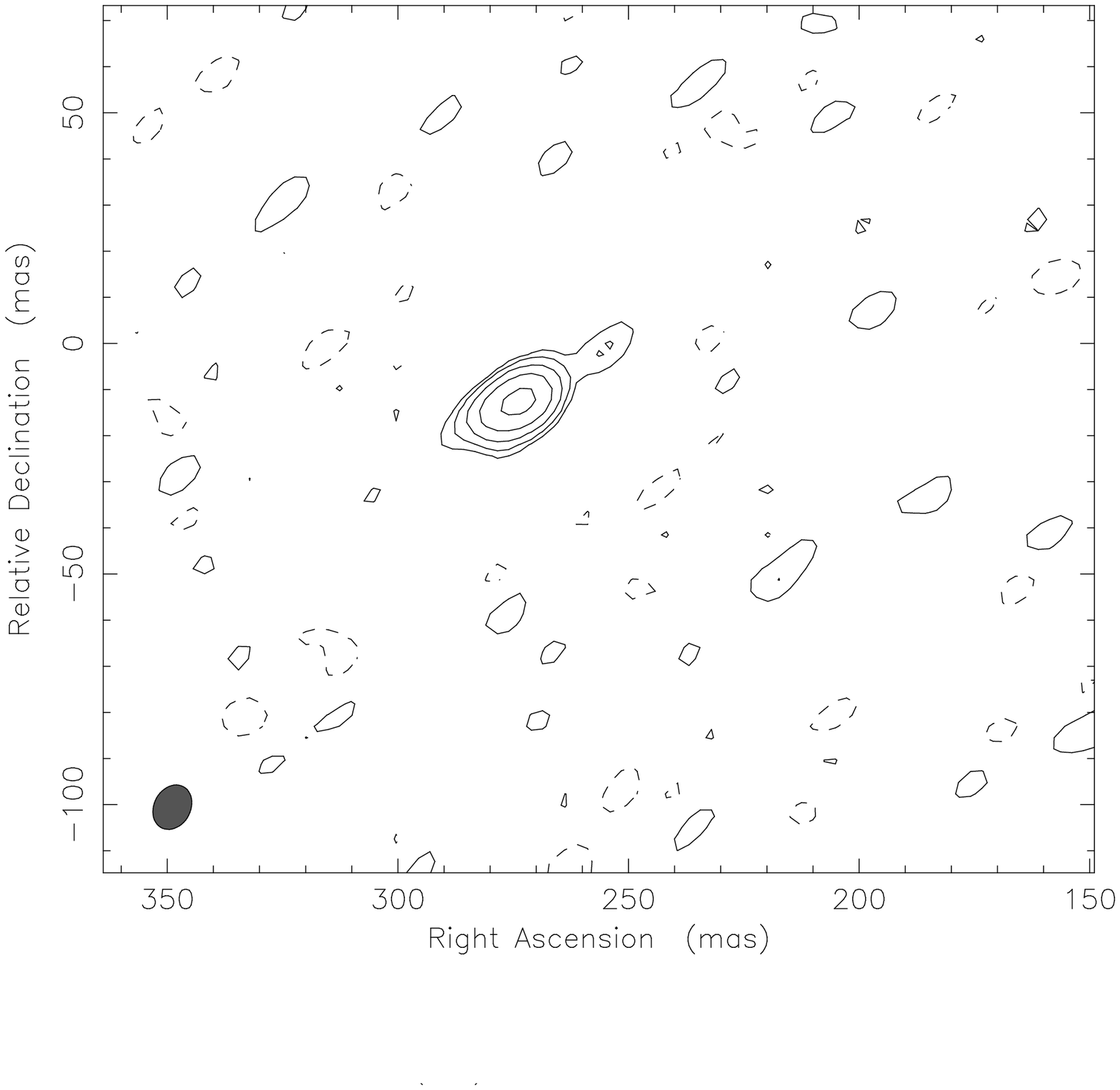}
\caption{(a: left) VLBA image of SDSS~J160024$+$264035 at 1.4 GHz.  The beam size is 7.4$\times$3.5 mas at a position angle of $-$28$^{\circ}$ and the contours are set at $-$50, $+$50, $+$70, and $+$90\% of the peak flux density of 1.5 mJy/beam. (b: right) VLBA image of SDSS~J151709$+$335324 at 1.4 GHz.  The beam size is 7.6$\times$3.2 mas at a position angle of $-$8$^{\circ}$ and the contours are set at $-$5, $+$5, $+$10, $+20$, $+40$ and $+$80\% of the peak flux density of 17 mJy/beam.}
\end{figure}

\clearpage

\begin{figure}
\plotone{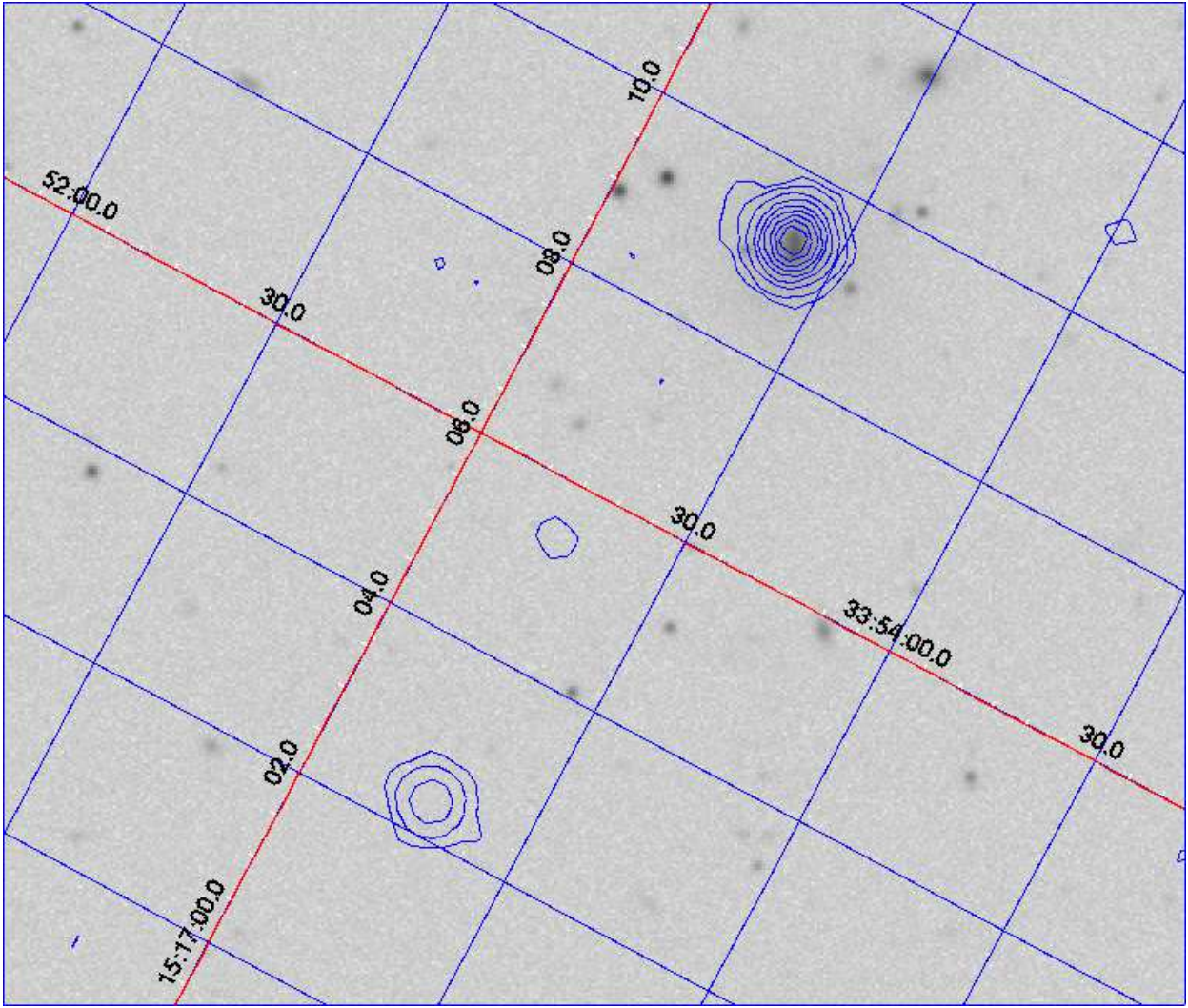}
\caption{FIRST and SDSS overlay image of SDSS~J151709$+$335324.  SDSS~J151709$+$335324 appears as the bright feature in the top right hand part of this figure.}
\end{figure}

\clearpage

\begin{deluxetable}{ccccrcc}
\tabletypesize{\scriptsize}
\tablecolumns{7}
\tablecaption{List of SDSS AGN observed with the VLBA}
\tablehead{
   \colhead{Source}                   &
   \colhead{RA (J2000)}               &
   \colhead{DEC (J2000)}              &
   \colhead{z}                        &
   \colhead{S$_{20 \rm{cm}}$ (mJy)}   &
   \colhead{S$_{\rm{VLBI}}$ (mJy)}    &
   \colhead{\arcsec/kpc}              
}
\startdata
SDSS J000249$+$004504& 00 02 49.081 &$+$00 45 04.92 &0.087 &2.33&$<$1.5&1.7 \\
SDSS J095833$-$005118& 09 58 33.247 &$-$00 51 18.74 &0.086 &1.63&$<$1.5&1.7 \\
SDSS J105653$+$331945& 10 56 53.384 &$+$33 19 45.12& 0.051 &2.53&$<$1.5&1.0 \\
SDSS J110957$+$020138& 11 09 57.130 &$+$02 01 38.50 &0.063 &6.37&$<$1.5&1.2 \\
SDSS J115249$+$190300& 11 52 49.309 &$+$19 03 00.50 &0.097 &2.37&$<$1.5&1.9 \\
SDSS J150452$+$321414& 15 04 52.343 &$+$32 14 14.96& 0.113 &1.67&$<$1.5&2.2 \\
SDSS J151659$+$051751& 15 16 59.253 &$+$05 17 51.43 &0.051 &5.84&$<$1.5&1.0 \\
SDSS J151709$+$335324& 15 17 09.289 &$+$33 53 23.92& 0.135 &120.30&17$\pm2$&2.6 \\
SDSS J152606$+$414014& 15 26 06.161 &$+$41 40 14.39 &0.008 &70.62&$<$1.5&0.1 \\
SDSS J155619$+$094855& 15 56 19.285 &$+$09 48 55.26 &0.068 &1.50&$<$1.5&1.3 \\
SDSS J160024$+$264035& 16 00 24.368 &$+$26 40 35.25 &0.090 &19.47&1.8$\pm0.2$&1.8 \\
\enddata
\label{tab:tabcomponents}
The source positions listed are the radio position from the FIRST catalog, accurate to approximately 0.5\arcsec.  The flux densities listed from FIRST are integrrated and have errors of approximately 10\%.  
\end{deluxetable}

\clearpage

\end{document}